\documentclass{article}%
\usepackage{amsmath}
\usepackage{graphicx}%
\usepackage{amsfonts}%
\usepackage{amssymb}

\begin{document}
\textbf{Specific Heat of URu}$_{2}$\textbf{Si}$_{2}\mathbf{\ }$\textbf{in
Fields to 42 T: \ Clues to the 'Hidden Order'}

\bigskip

J. S. Kim, D. Hall, P. Kumar, and G. R. Stewart

Department of Physics, University of Florida, Gainesville, FL \ 32611-8440

\bigskip

Abstract: \ The large $\Delta$C observed at 17.5 K in URu$_{2}$Si$_{2}$\ is
inconsistent with the small, 0.04 $\mu_{B}$ moment measured for the
antiferromagnetism observed starting (perhaps coincidentally) at the same
temperature. \ We report measurements of this specific heat transition,
thought to be due to some 'hidden order', in magnetic fields between 24 and 42
T, i. e. through the field region where three metamagnetic transtions are
known to occur at 35.8, 37.3, and 39.4 T. \ The response of $\Delta$C in
single crystal URu$_{2}$Si$_{2}$ to magnetic field, which includes a change to
$\Delta$C\ being possibly associated with a first order phase transition for
high fields, is analyzed to shed further light on the possible explanations of
this unknown ordering process. \ \ At fields above 35 T, a new high field
phase comes into being; the connection between this high field phase revealed
by the specific heat and earlier magnetization data is discussed.

\bigskip

\pagebreak 

I. \ Introduction

\bigskip

\qquad The compound URu$_{2}$Si$_{2}$\ was initially a focus of research due
to the discovery$^{1}$ of the coexistence of antiferromagnetism (T$_{N}$ =
17.5 K) and superconductivity. Later work$^{2-3}$ focussed on the three steps
in the magnetization (called 'metamagnetic transitions') discovered$^{4}$ by
magnetoresistance and magnetization measurements at 35.8, 37.3, and 39.4 T
with the field parallel to the c-axis in this medium heavy fermion compound .
\ Recently, the explanation for the large (in proportion to the ordered
moment$^{5}$ of only 0.04 $\mu_{B}$) size of the discontinuity in the specific
heat at 17.5 K in URu$_{2}$Si$_{2}$ has been the subject of much theoretical
as well as experimental effort$^{6-22}$, with explanations as diverse as
quadrupolar order$^{16}$ and various unusual kinds of magnetic order being
proposed$^{23}$. \ \ (Some of these latter have been found to be inconsistent
with later neutron diffraction experiments$^{8}$.) \ The term 'hidden order'
has been coined to refer to the as yet unknown order that leads to the large
observed $\Delta$C, i. e. 'hidden' in the sense that the nature of the order
has not yet revealed itself to a variety (e. g. neutron scattering$^{8}$,
resistivity$^{9}$, and NMR$^{17}$) of measurement techniques that have been
used to probe the order responsible for the large $\Delta$C.

\qquad It is apparent from various perspectives that the observed anomaly in
the specific heat is \textit{not} due to the observed$^{8}$ low moment
magnetism. \ One argument (see ref. 10) is that Landau theory predicts that
$\Delta$C/T $\approx$ (k$_{B}$/T$_{N}$) times the square of the ratio of the
observed moment to the paramagnetic moment observed (either from neutron
scattering or from an analysis of the magnetic susceptibility fitted to the
Curie-Weiss law) above T$_{N}$. \ For URu$_{2}$Si$_{2}$, this gives$^{10}$ a
prediction for $\Delta$C/T that is three orders of magnitude smaller than that
observed. \ \ Recent neutron diffraction$^{19}$ and NMR$^{20}$ measurements
under pressure have been interpreted to imply that the observed magnetic
moment is in fact from a minority (\symbol{126}1\%) second phase, although
this is still under discussion$^{22}$.

\qquad Hall effect data to 40 T by Bakker, et al.$^{3}$ show transitions in
the Hall coefficient at 35.6, 36.2, and 39.2 T (i. e. the Hall effect data
appear linked to the three transitions at 35.8, 37.3, and 39.4 T in the
magnetization data). \ The decrease of the Hall coefficient down to a value
close to zero at 39.2 T was interpreted$^{3}$ as a closing of the gap in the
Fermi surface which was opened at 17.4 K. \ This energy gap of about 110 K,
measured below the ordering temperature in zero and applied field by various
methods$^{9}$, was seen by resistivity data$^{9}$ in fields to 25 T to be
associated with order that had a critical field of 40 T, i.e. with the
'hidden' order. \ Thus, since the Hall effect data appear to show the closing
of the gap at 40 T associated$^{9}$ with the hidden order and since the Hall
effect appears also to have three transitions comparable to those shown in the
magnetization, perhaps the 'metamagnetic' transitions in M vs H between 35.8
and 39.4 T are also associated with the hidden order.

\qquad Since the deciding measurement that determines the existence of the
'hidden' order is the specific heat, since the specific heat (resistivity) of
URu$_{2}$Si$_{2}$ has been reported$^{12}$ only in fields up to 17.5 (25$^{9}%
$) T, and since there is clearly something unusual occurring in the phase
diagram as a function of magnetic field at 35.8, 37.3, and 39.4 T, we
undertook to measure$^{24}$ specific heat for the first time in dc fields up
to 42 T in the new dc hybrid magnet at the NHMFL in Tallahassee on single
crystals of URu$_{2}$Si$_{2}$. \ (Prior to the advent of this unique magnet,
all measurements above 35 T had to be performed in pulsed field magnets.)
\ These measurements should help to further elucidate the nature of the
'hidden' order by determining the response of the anomaly over the entire
field range expected$^{9}$ to suppress the anomaly and also by determining how
the anomaly reacts (if at all) at the fields where the jumps in the
magnetization occur. \ 

\bigskip

\bigskip II. \ Experimental

\bigskip\qquad Although there are some minor differences$^{25}$ in measuring
specific heat in fields above 24 T versus techniques used$^{24}$ for H$\leq$24
T, in general the measurements are quite similar. \ In order to avoid eddy
current heating (much more a problem at the old Francis Bitter National Magnet
Laboratory than in the new National High Magetic Field Laboratory due to the
quieter, modern transistor based power supply used at NHMFL), the reference
block$^{24}$ used is made out of non-electrically conducting sapphire. \ The
transfer standard to calibrate the sample platform thermometer (a
flash-evaporated film of Au-Ge with$^{26}$ a \textit{positive}
magnetoresistance - i. e. magnetic field enhances the thermometer's
sensitivity - of about 20\% at 1 K in 33 T) is a capacitance thermometer from
Lakeshore Cryotronics that is essentially$^{27}$ field independent up to 45 T.\qquad

\qquad Flat platelet single crystals, with the c-axis perpendicular to the
platelet, were obtained by removing the crystals that form as surface facets
on a large (\symbol{126}2 g mass) arc-melted button of high purity URu$_{2}%
$Si$_{2}$, a technique that has also been used$^{28}$ to produce single
crystals of CeRu$_{2}$Si$_{2}$. \ A collage of 16.18 mg of these crystals
thermally bonded to a sapphire disk using GE7031 varnish was used for the
specific heat measurements. \ The Ru and Si used were 99.95\% and 99.9999\%
pure respectively from Johnson Matthey; the U used was the best that is
commercially available, electrotransport refined material from Ames Lab. \ The
zero field specific heat of these crystals is comparable with the best samples
reported, with $\Delta$C/T$_{order}$ = 335 mJ/molK$^{2}$ at the high
temperature ordering transition in the present work, vs 320 mJ/molK$^{2}$ in
ref. 12. ($\Delta$C is defined as C/T$_{\max}$-C/T extrapolated from the
higher temperature, normal state data down to T$_{\max}$.) \ Further, the
transition width of the hidden order phase transition in zero field in the
sample used in the present work is 0.30 K vs 0.38 K in the sample measured in
ref. 12. \ The susceptibility with field parallel to both the c- and a-axes
was also measured (where the two directions have a factor of \symbol{126}eight
difference in \ magnitude as well as a much different temperature dependence),
with excellent agreement with the literature$^{1}$ results. \ 

\bigskip

III. \ Results and Discussion

\bigskip

\qquad The specific heat divided by temperature, C/T, of single crystal
URu$_{2}$Si$_{2}$ between 24 and 35 T (where the first metamagnetic transition
is reported at 35.8 T) is plotted in Fig. 1, between 35 and 37.5 T in Fig. 2,
and between 37.5 and 42 T in Fig. 3. \ The discussion of these data divides
naturally into two field regions.%
\begin{figure}
[ptb]
\begin{center}
\includegraphics[
natheight=2.927400in,
natwidth=3.596800in,
height=2.4102in,
width=2.9568in
]%
{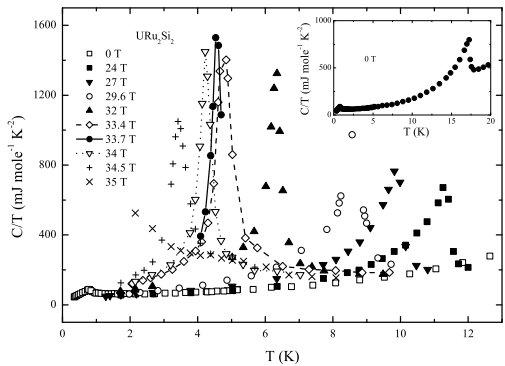}%
\caption{{\small Specific heat, C, divided by temperature, T, at low
temperatures between 0 and 35 T for single crystal URu}$_{2}${\small Si}$_{2}%
${\small with the field aligned along the c-axis. \ Note the very rapid change
of both the magnitude of }$\Delta${\small C/T}$_{order}${\small and
T}$_{order}${\small for fields above 33.7 T: \ in only 1.3 T T}$_{order}%
${\small decreases more than 2.5 K as the hidden order phase transition is
being finally suppressed to T=0.\ \ }}%
\end{center}
\end{figure}

\bigskip

A. \ H $\leq$ 35 T - Suppression of the Hidden Order Anomaly

\bigskip

\qquad\ \ The trend of the ordering temperature decreasing and $\Delta
$C/T$_{order}$ increasing with increasing field up to 33.7 T apparent in Fig.
1 was also qualitatively observed by van Dijk, et al.$^{12\text{ }}$in their
specific heat data in fields up to 17. 5 T. \ \ As may be seen, the anomaly in
C/T sharpens up remarkably with increasing field above 27 T, becoming in
appearance (see discussion below) like a first order phase transition. \ This
sharpness of the anomaly makes the accurate determination of T$_{order}$ for
each field straightforward. \ 

\qquad One result of these measurements is the critical field of the specific
heat anomaly as a function of temperature, shown in Fig. 4. \ The temperature
dependence of the critical field of this 'hidden' order anomaly up to 35 T is
not unusual at all, and follows at low field (as was known from, e. g., the
specific heat data up to 17.5 T from ref. 12) H$_{c}$ = H$_{0}$%
[1-(T/T$_{order}$)]$^{0.5}$ rather well, where H$_{0}$\symbol{126}35.3 T and
T$_{order}$ is the zero field ordering temperature (defined as the temperature
of the peak in C/T, rather%
\begin{figure}
[ptb]
\begin{center}
\includegraphics[
natheight=7.254900in,
natwidth=9.000100in,
height=2.1032in,
width=2.6057in
]%
{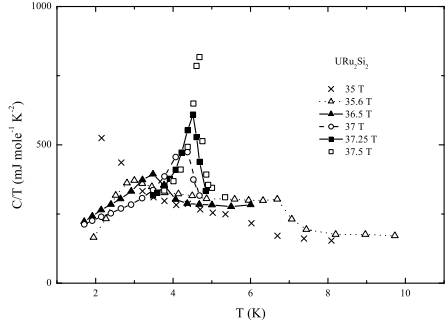}%
\caption{{\small C/T between 35 and 37.5 T for single crystal URu}$_{2}%
${\small Si}$_{2}${\small with H}$\parallel${\small c-axis. \ Although any
peak in 35 T occurs below 2 K, already by 35.6 K a small anomaly is visible in
C/T near 3 K which, as discussed in the text, is the appearance of a high
field phase. T}$_{order}${\small and }$\Delta${\small C/T}$_{order}%
${\small for this anomaly then continue to increase with increasing field in
the field range shown here. \ A second anomaly at higher temperatures
(\symbol{126}6 K) may be apparent in the 35 T data but is clearly visible in
the 35.6 T data. \ Data in this temperature range were not taken again until
38.1 T (at which point the small anomaly appears to be absent, see Fig. 3).
\ This anomaly may correspond to a second high field anomaly bounded by the
two lower dashed lines shown in Fig. 4 corresponding to the two lower
metamagnetic transitions at 35.8 and 37.3 T.}}%
\end{center}
\end{figure}
\begin{figure}
[ptbptb]
\begin{center}
\includegraphics[
natheight=7.299000in,
natwidth=9.000100in,
height=2.1421in,
width=2.6377in
]%
{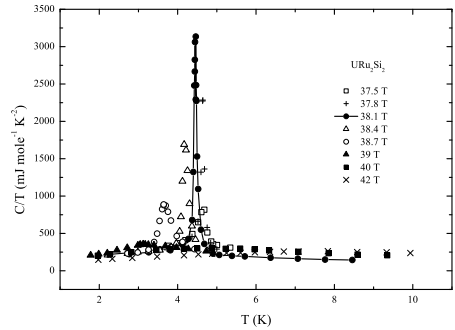}%
\caption{{\small C/T between 37.5 and 42 T for single crystal URu}$_{2}%
${\small Si}$_{2}${\small with the field aligned along the c-axis.
\ T}$_{order}${\small decreases monotonically for fields starting at 37.5 T,
and }$\Delta${\small C/T}$_{order}${\small has a maximum at 38.1 T.}}%
\end{center}
\end{figure}
than the onset of the transition) of 17.4 K in URu$_{2}$Si$_{2}$. \ At high
field, i. e. near H$_{0}$=35.3 T, our data shown in Fig. 4 allow the
determination that H$_{c}$ follows H$_{0}$[1-(T/T$_{order}$)$^{2}$]. \ \ These
two temperature dependences in the two limits T$\rightarrow$T$_{order}$ and
T$\rightarrow$0 provide useful information for theories (e. g. ref. 7) which
examine the nature of the hidden order.%
\begin{figure}
[ptbptbptb]
\begin{center}
\includegraphics[
height=3.3503in,
width=2.5624in
]%
{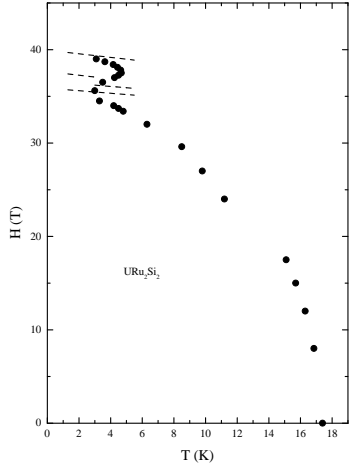}%
\caption{{\small The critical field is plotted versus T}$_{order}${\small for
single crystal URu}$_{2}${\small Si}$_{2}${\small with the field aligned along
the c-axis using the temperature of the maximum in C/T for T}$_{order}%
${\small . \ The data for H}$\leq${\small 17.5 T are from ref. 12, while the
dashed lines at high fields - as discussed in the text - represent the
temperature dependence of the fields where the three jumps in the
magnetization occur from ref. 24. \ Note the rapid approach to a zero slope
for dH/dT as H}$\rightarrow${\small 35 T, as well as the high field phase that
is induced by field between 35.6 and 39 T (data from Figs. 2 and 3.)}}%
\end{center}
\end{figure}

\qquad Considering now the field behavior of $\Delta$C/T$_{order}$, this is
plotted in Fig. 5. \ Special care was taken while measuring the data to
accurately determine the field where the maximum in C/T occurs. \ As clear
from Figs. 1 and 5, this field is 33.7 T to an accuracy of better than 0.3 T.
\ This does not agree with the field determined in magnetization measurements
for the first jump in M as a function of field, which is at 35.8 T. \ \ This
disagreement is also apparent when considering the entropy as a function of
field (also shown in Fig. 5). \ These facts rule out explanations for the jump
in magnetization (related to the entropy through the Clausius-Clapeyron
equation) which involve large changes in the entropy (e. g. level crossing
transitions). \ It is interesting to note that the Hall effect data$^{3}$
taken up to 40 T show an anomaly (unremarked upon in ref. 3) at 0.6 K between
33.8 and 34.5 T (i. e. similar to the field of 33.7 T where we see the maximum
in C/T) that is actually slightly \textit{larger} than the anomaly between
35.1 and 36.1 T that was identified$^{3}$ as corresponding to the
magnetization anomaly at 35.8 T.%
\begin{figure}
[ptb]
\begin{center}
\includegraphics[
natheight=2.895400in,
natwidth=3.966900in,
height=1.4754in,
width=2.0167in
]%
{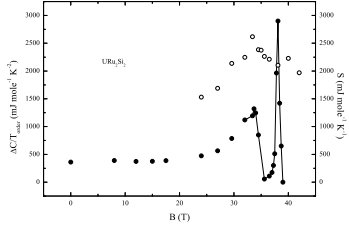}%
\caption{$\Delta${\small C/T}$_{order}${\small for the anomalies in the
specific heat (shown in Figs. 1-3) plotted as a function of field (data for
H}$\leq${\small 17.5 T from ref. 12). \ These values have not been corrected
for the jump due to the first order phase transition and thus overestimate the
quantity }$\Delta${\small C/T in the theoretical treatment in ref. 7. \ Note
the two peaks in }$\Delta${\small C/T}$_{order}${\small at 33.7 and 38.1 T,
with the higher field peak occurring over a very narrow range of field.
\ \ Also shown is the entropy up to 10 K as a function of field, showing
(although with a less fine resolution than for }$\Delta${\small C/T}$_{order}%
${\small , since data up to 10 K necessary to calculate S(10 K) were not taken
for every field due to time constraints) that also the peak in the entropy
clearly occurs at a field below that of the jump in the magnetization. \ In
relating via dS/dH = dM/dT these entropy results to what is known about M as a
function of field and temperature, the relation d}$^{2}${\small S/dH}$^{2}%
${\small = d/dT (dM/dH) implies - since we know that the jump in the
magnetization with increasing field at 35.8T broadens with increasing
temperature - that d}$^{2}${\small S/dH}$^{2}${\small should be negative
around 35.8 T. \ With the accuracy of the entropy data and the spacing of the
data as a function of field, this relation between second derivatives is
difficult to confirm.}}%
\end{center}
\end{figure}
\begin{figure}
[ptbptb]
\begin{center}
\includegraphics[
natheight=2.927400in,
natwidth=3.596800in,
height=3.1842in,
width=3.9046in
]%
{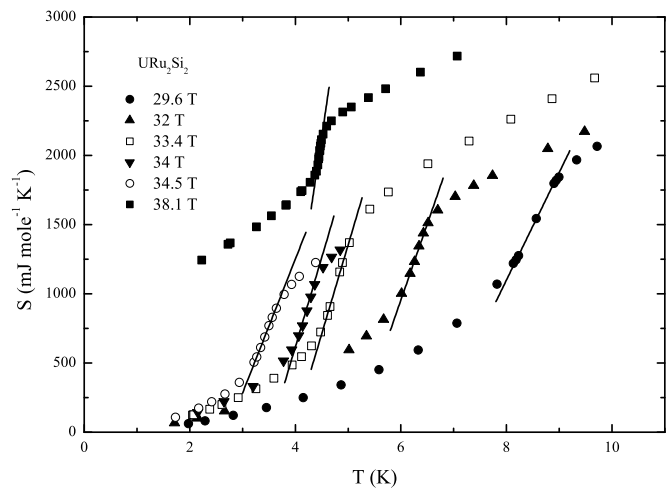}%
\caption{{\small Entropy, S, vs temperature for single crystal URu}$_{2}%
${\small Si}$_{2}${\small with the field aligned along the c-axis for fields
between 29.6 and 34.5 T, as well as for 38.1 T (plotted with the same vertical
and horizontal scales but shifted upwards by 1500 mJ/moleK to make the data
more visible). The entropy (}$\equiv$$\int_{0}^{T}${\small C/T' dT') is
calculated numerically from the C/T data shown in Fig. 1 and, for 38.1 T, in
Fig. 3. \ For a second order phase transition, or for a first order phase
transition rounded by local impurities or defects}$^{29}${\small , only one
inflection point is expected in the S vs T curve. \ Note the tendency for the
slope of S vs T to become steeper around 34 T. \ Note also that the slope of S
vs T for the 38.1 T data is much higher (factor of two) than for any other
field, i. e. the field-induced transition at 38.1 T appears to be the most
likely to be first order in nature.}}%
\end{center}
\end{figure}

\qquad The sharpness of the anomalies shown in Fig. 1 for H $\geq$ 29.6 T
raises the questions of whether the hidden order anomaly in $\Delta$C/T
becomes first order in high field, and\textit{\ }if so, at what field this
first occurs. \ If the hidden order phase transition becomes first order with
increasing magnetic field, this would certainly be an important constraint for
any theory explaining this as-yet not understood order. \ To address this
question, plots of the entropy, S, for fields $\geq$ 29.6 T\ are presented in
Fig. 6. \ \ Leaving discussion of the 38.1 T entropy results also shown in
Fig. 6 for the next section, it is clear from Fig. 6 that, although the slope
of S vs T is a maximum around 34 T and despite the quite large and sharp
appearing specific heat anomalies in Fig. 1, the entropies never have the
truly vertical jump characteristic of a first order phase transition. \ \ In
order to address the question of the possibility that inhomogeneities in the
collage of single crystals created a broadening of the jump in the entropy,
and to improve the data density vs temperature (see, e. g., Fig. 1) achieved
within the (severe) time constraints of measuring in the 45 T hybrid magnet,
we have measured during a separate week at NHMFL the specific heat of the
largest URu$_{2}$Si$_{2}$ crystal%
\begin{figure}
[ptb]
\begin{center}
\includegraphics[
natheight=2.704300in,
natwidth=3.596800in,
height=2.2286in,
width=2.9568in
]%
{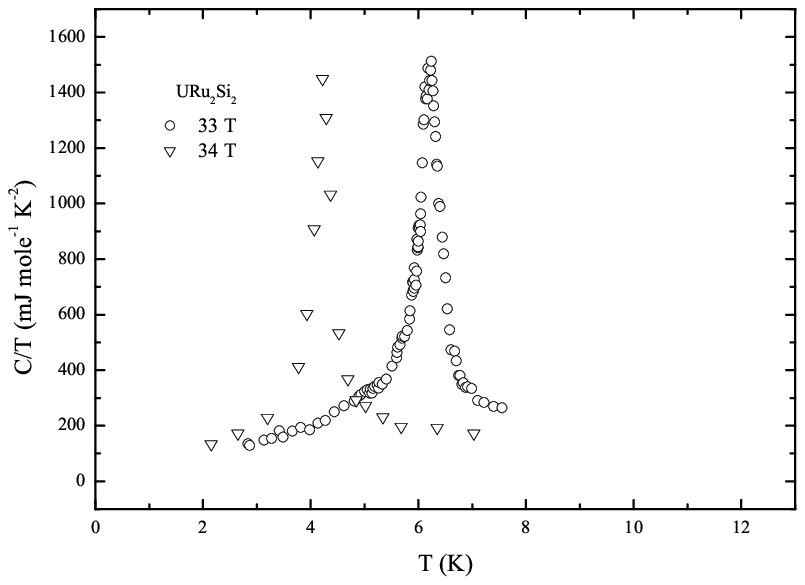}%
\caption{{\small C/T vs T for a 5.03 mg single crystal of URu}$_{2}%
${\small Si}$_{2}${\small with the field aligned along the c-axis in 33 T
(high density data as discussed in the text to address the order of the
transition) and, for comparison, data in 34 T from Fig. 1. \ These 33 T data
were taken with much smaller (0.01 K vs 0.05 K) temperature excursions}$^{24}%
${\small than for the other data in this work\ in order to smear out the
transition as little as possible.}}%
\end{center}
\end{figure}
\begin{figure}
[ptbptb]
\begin{center}
\includegraphics[
natheight=2.900600in,
natwidth=3.654700in,
height=2.3307in,
width=2.9308in
]%
{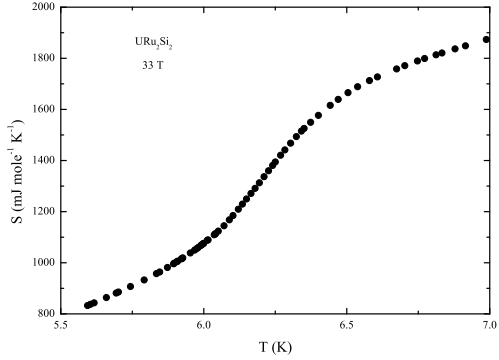}%
\caption{{\small Entropy, S, vs temperature for single crystal URu}$_{2}%
${\small Si}$_{2}${\small with the field aligned along the c-axis for 33 T
using the data from Fig. 7.}}%
\end{center}
\end{figure}
(5.03 mg) from the collage in 33 T (see Fig. 7). \ These measurements were
performed using the most painstaking data density possible with our
measurement technique, with very small temperature excursions and a very high
density of measurement temperatures around the peak in C/T. \ \ (Measurement
time in a magnet that reaches 33 T is easier to obtain than time in the
unique, 45 T hybrid magnet at NHMFL.) \ \ The entropy calculated from these
high density data in 33 T is shown in Fig. 8. \ The slope of the steepest part
of the entropy vs temperature data of this single crystal data with much
higher data density, Fig. 8, is only about 9 \% higher than that of the data
for the most comparable field, 33.4 T, data shown in Fig. 6. \ Thus, unless
there is a high level (significantly above what is expected) of either a.)
internal (e. g. variation in stoichiometry) inhomogeneities in the individual
single crystal, b.) inhomogeneities in the field$^{30}$, or c.) impurities or
defects$^{29}$ in the single crystal, the hidden order transition for fields
$\leq$ 34.5 T appears to be second order.

\qquad Above 33.7 T (see Figs. 1 and 5), the size of the anomaly in C/T falls
precipitously, well before the first anomaly in the magnetization at 35.8 T.
\ \ At 35 T, the peak in the C/T anomaly is below our lowest temperature of
measurement (2 K) \ and is (based on an extrapolation of the data at 34 and
34.5 T where the peak is still visible) smaller by about a factor of three
from the maximum (see Fig. 5) in $\Delta$C/T$_{order}$ at 33.7 T. \ One
possible explanation for the rather precipitous fall in T$_{order}$ and the
size of the transition between 34.5 and 35 T shown in Fig. 1 is best seen by
considering the phase diagram shown in Fig. 4. \ As the hidden order
transition temperature is suppressed with increasing field towards lower
temperatures, Fig. 4 shows that this phase line becomes more and more
horizontal as H$\rightarrow$35 T(or slightly greater). \ At this point in a
phase diagram, measurements in a given field as a function of temperature will
show a broadening of any field dependent phase transition. \ Measurements to
lower temperatures would help further investigate this possibility.

\bigskip

B. \ 35 T $\leq$ H $\leq$ 42 T - A New High Field Phase

\bigskip

\qquad The specific heat data between 35 and 37.5 T \ and between 37.5 and 42
T are shown in Figs. 2 and 3 respectively. \ A specific heat anomaly at
\symbol{126}3 K is seen to be moving up in temperature with increasing field
already in 35.6 T after the hidden order anomaly was suppressed below 2 K in
35 T. \ While the second anomaly in the magnetization data taken at 1.3 K is
at$^{4}$ 37.3 T (or 37.4 T in the more recent work of Sugiyama, et al.$^{31}%
$), the data in Fig. 2 show a monotonic increase in both the size of $\Delta
$C/T$_{order}$ as well as T$_{order}$ up to 37.5 T, after which (see Fig. 3)
the anomaly continues to increase in size up to 38.1 T, but T$_{order}$ begins
to decrease for H
$>$
37.5 T. \ The fields and T$_{order}$ are shown in Fig. 4; clearly the high
field phase diagram shows another phase transition that appears after the
hidden order anomaly is suppressed around 35 T. \ Considering now the entropy,
S(10 K) shown in Fig. 5 trends slightly downwards in this field range after
its peak around 33 T.

\qquad A second, small anomaly at \symbol{126}6.5 K is visible in the 35.6 T
data, with indications of a related anomaly in the trend of the 36.5 T data up
to 6 K. \ This smaller anomaly is absent by 38.1 T, and - as discussed below -
\underline{may} correspond to a small second field-induced transition in
addition to the very large anomaly in C/T with a maximum as a function of
field at 38.1 T. \ Further data in the field and temperature ranges 35 - 38 T
and 4 - 10 K are needed to resolve this. \ Another possibility is based on the
discussion of the phase diagram and the 35 T data at the end of the section
above: \ this broadened anomaly in 35.6 T is again where a phase boundary line
is nearly horizontal. \ Thus, the broadened anomaly with its beginning at
\symbol{126}6.5 K may in fact just be due to the spreading out of a phase
transition as the phase boundary for the new, higher field phase joins the
field axis horizontally. \ Clearly from Fig. 2, the anomaly above 36.5 T
begins to sharpen and increase in temperature, corresponding to a finite slope
for the phase boundary at increasing fields as shown in Fig. 4.

\qquad Sugiyama, et al.$^{31}$ consider the temperature dependence of the
three magnetization anomalies (shown as dashed lines in Fig. 4), which
addresses the intercomparability of the anomalies at magnetization measured -
in most cases - at different points (1.5 K and 35.8, 37.3, and 39.4 T in the
early work of de Visser, et al.$^{4}$ and at 1.3 K and 35.4, 37.4, and 39.2 T
in the more recent work of Sugiyama, et al.$^{31}$) in the H,T phase diagram
than the anomalies in C/T. \ They find that the anomaly in the magnetization
at 1.3 K and 37.4 T disappears by around 3 K and a new anomaly appears at 36 T
faintly in their 3 K data but grows in size and is quite visible in their 3.6
and 4.2 K data. \ This anomaly is comparable to the part of the H,T phase
space sampled by our specific heat data in 36.5 T, where an anomaly in C/T is
peaked at 3.5 K. \ \qquad

\qquad Considering now the specific heat data in the field range around the
third anomaly in the magnetization, Fig. 3 (see Fig. 4 for the resultant phase
diagram), we see that the growing anomaly in C/T evident in 37.5 T in Fig. 2
rises up very sharply as a function of field (see also Fig. 5), peaking in
magnitude at 38.1 T and 4.44 K in a transition that - even more than the
transitions shown in Fig. 1 - is reminiscent of a first order phase
transition. \ One important comparison (see Fig. 6) is the steepness of the S
vs T curve at the anomaly. \ As shown in Fig. 6, the anomaly at 38.1 T is
clearly, based on the criterion of how steeply S rises as a function of
temperature, more first order in appearance. \ (Quantitatively, the slopes of
the entropies at 38.1 and 33.4 T at their steepest sections differ by a factor
of two.) \ \ The question then arises if the transition evinced by the (rather
steep) entropy anomaly in 38.1 T shown in Fig. 6 is a broadened first order
phase transition, or simply a rather sharp second order phase
transition.\ \ Thus, just as a second order phase transition should
theoretically have a discontinous change in the specific heat at the
transition - but in fact there is always a finite and sometimes a significant
transition width $\Delta$T in the specific heat jump, so may a first order
phase transition have a finite transition width $\Delta$T in the entropy jump.
\ In zero field at 17.5 K, our sample of single crystal URu$_{2}$Si$_{2}$ has
sufficient inhomogeneity to have a $\Delta$T of 0.30 K in its specific heat
jump. \ The width of the steep increase in the entropy at the transition in
38.1 T is less than 0.20 K. \ High data density, low temperature excursion
data will be taken on the 5.03 mg single crystal of URu$_{2}$Si$_{2}$ just as
done for the 33 T data discussed above as soon as hybrid magnet time is
available in order to further investigate the possible first order nature of
the 38.1 T transition.

\qquad In the phase diagram constructed by Sugiyama, et al.$^{31}$, the
highest field anomaly occurs at 39.1 T for 4.4 K. \ This disagreement, plus
the complete lack of any hint in the magnetization data of a rapid variation
with field of the magnitude of this third anomaly imply once again that the
specific heat results are presenting new information about the phase diagram
of URu$_{2}$Si$_{2}$ in high magnetic field. \ In the specific heat measured
at various fields as a function of temperature, an anomaly appears in C/T at
35.6 T and grows smoothly in magnitude (see Fig. 5) with increasing field up
to 38.1 T, with a peak in T$_{order}$ of 4.68 K at the slightly lower field of
37.5 T. \ This is incorporated in the high field portion of the phase diagram
shown in Fig. 4. \ 

\qquad The magnetization, rather than showing a region of existence in the
phase diagram of the high field phase, shows$^{31}$ a smooth, gradual decrease
in the fields of the upper and lower anomalies with increasing temperature,
and a decrease in the field of the middle anomaly with increasing temperature
that shows a sudden decrease at around 3 K. \ Thus, the
magnetization-data-generated phase diagram shows the three separate,
apparently independent dotted lines drawn in Fig. 4. \ Comparing the
magnetization and the specific heat data graphically in the phase diagram in
Fig. 4 suggests that the two higher field magnetization anomalies may mark the
\textit{boundaries} of the large $\Delta$C/T\ high field phase transtion
observed in the specific heat. \ Whether or not the slight anomaly in the 35.6
T data in Fig. 2 corresponds to a second high field phase that is bounded by
the lower two of the dashed lines in Fig. 4 representing the H,T behavior of
the two magnetization anomalies at 35.8 and 37.3 T depends on the outcome of
further work. \ It is interesting to note the possible connection in the jump
in the magnetization phase line for the middle magnetization anomaly and the
swerving away from joining the ordinate for the lower boundary of the specific
heat phase boundary shown in Fig. 4.

\bigskip\ 

IV. Conclusions

\bigskip

\qquad Specific heat indicates that the hidden order anomaly is suppressed at
\symbol{126}35 T, and that the hidden order anomaly in high field before its
suppression appears to remain second order. \ \ The order of this transition
is important for understanding the nature of the hidden order in URu$_{2}%
$Si$_{2}$ and needs to be pursued in further work since whether the anomaly in
the specific heat for H$\cong$33.7 T becomes first order is only approximately
determined by the data in the present work. \ In addition, the specific heat
data reported here show for the first time the existence of a high field phase
in URu$_{2}$Si$_{2}$ which is apparently linked to the two higher field
magnetization anomalies first seen in 1987$^{4}$. \ This new high field phase
may be first order in nature. \ Indications of a second high field phase
observed in the narrow field range between 35.6 and 38.1 T may also be
explained by the rather flat H vs T phase boundary in this part of the phase diagram.

\bigskip

\textit{Note Added in Proof: \ }Recently, M. Jaime et al.$^{32}$ have measured
the high field specific heat of a single crystal sample of URu$_{2}$Si$_{2}$
in dc fields up to 45 T, with a spacing of approximately 2 T between
measurement fields in the field range between 30 and 40 T. \ They also find
the new high field phase reported herein and, using magnetocaloric
measurements in pulsed fields, also find evidence that suggests that the
transition at \symbol{126}39 T is ''a first order-like transition in field.''

\bigskip

Acknowledgements: \ The authors gratefully acknowledge helpful discussions
with Prof. M. Wortis. \ Work at the University of Florida performed under the
auspices of the U. S. Department of Energy, contract \#DE-FG05-86ER45268.
\ Data were taken at the NHMFL, Tallahassee, which is operated under the
auspices of the U. S. National Science Foundation.

\pagebreak 

References

\bigskip

1. \ W. Schlabitz, J. Baumann, B. Pollit, U. Rauchschwalbe, H. M. Mayer, U.
Ahlheim, and C. D. Bredl, Z. Phys. B \textbf{62, }171 (1986); T. T. M.
Palstra, A. A. Menovsky, J. van den Berg, A. J. Dirkmaat, P. H. Kes, G. J.
Nieuwenhuys, and J. A. Mydosh, Phys. Rev. Lett. \textbf{55}, 2727 (1985); \ M.
B. Maple, J. W. Chen, Y. Dalichaouch, T. Kohara, C. Rossel, M. S.
Toricachvili, M. W. McElfresh, and J. D. Thompson, Phys. Rev. Lett.
\textbf{56}, 185 (1986).

2. \ K. Sugiyama, H. Fuke, K. Kindo, K. Shimohata, A. A. Menovsky, J. A.
Mydosh, and M. Date, J. Phys. Soc. Japan \textbf{59}, 3331 (1990).

3. \ K. Bakker, A. de Visser, A. A. Menovsky, and J. J. M. Franse, Physica B
\textbf{186-188}, 720 (1993).

4. A. de Visser, F. R. deBoer, A. A. Menovsky, and J. J. M. Franse, Solid
State Commun. \textbf{64}, 527 (1987).

5. C. Broholm, H. Lin, P. T. Matthews, T. E. Mason, W. J. L. Buyers, M. F.
Collins, A. A. Menovsky, J. A. Mydosh, and J. K. Kjems, Phys. Rev.
B\textbf{43}, 12809 (1991).

6. P. Chandra, P. Coleman, and J. A. Mydosh, Physica B, in press.

7. N. Shah, P. Chandra, P. Coleman, and J. A. Mydosh, Phys. Rev. B\textbf{61},
564 (2000).

8. T. E. Mason, W. J. L. Buyers, T. Petersen, A. A. Menovsky, and J. D.
Garrett, J. Phys.: \ Condens. Matter \textbf{7}, 5089 (1995).

9. S. A. M. Mentink, T. E. Mason, S. Suellow, G. J. Nieuwenhuys, A. A.
Menovsky, J. A. Mydosh, and J. A. A. J. Perenboom, Phys. Rev. B\textbf{53},
R6014 (1996).

\bigskip10. W. J. L. Buyers, Physica B\textbf{223\&224}, 9 (1996).

11. S. A. M. Mentink, U. Wyder, J. A. A. J. Perenboom, A. de Visser, A. A.
Menovsky, G. J. Nieuwenhuys, J. A. Mydosh, and T. E. Mason, Physica
B\textbf{230-232}, 74 (1997).

12. N. H. van Dijk, F. Bourdarot, J. C. P. Klaasse, I. H. Hagmusa, E. Brueck,
and A. A. Menovsky, Phys. Rev. B\textbf{56}, 14493 (1997).

13. L. P. Gor'kov and A. Sokol, Phys. Rev. Lett. \textbf{69}, 2586 (1992).

14. Y. Miyako, S. Karawarazaki, H. Amitsuka, C. C. Paulsen, and K. Hasselbach,
J. Appl. Phys. \textbf{70}, 5791 (1991).

15. A. P. Ramirez, P. Coleman, P. Chandra, E. Brueck, A. A. Menovsky, Z. Fisk,
and E. Bucher, Phys. Rev. Lett. \textbf{68}, 2680 (1992).

16. P. Santini, and G. Amoretti, Phys. Rev. Lett. \textbf{73}, 1027 (1994).

17. O. O. Bernal, C. Rodrigues, A. Martinez, H. G. Lukefahr, D. E.
MacLaughlin, A. A. Menovsky, and J. A. Mydosh, Phys. Rev. Lett. \textbf{87},
196402 (2001).

18. P. Chandra, P. Coleman, J. A. Mydosh, and V. Tripathi, ''Hidden Orbital
Order in URu$_{2}$Si$_{2}$'', Nature \textbf{417}, 831 (2002).

19. \ H. Amitsuka, N. Sato, N. Metoki, M. Yokohama, K. Kuwahara, T.
Sakakibara, H. Morimoto, S. Kawarazaki, Y. Miyako, and J. A. Mydosh, Phys.
Rev. Lett. \textbf{83}, 5114 (1999).

20. \ K. Matsuda, Y. Kohori, T. Kohara, K. Kuwahara, and H. Amitsuka, Phys.
Rev. Lett. \textbf{87}, 087203 (2001).

21. \ H. Amitsuka, M. Yokoyama, S. Miyazaki, K. Tenya, T. Sakakibara, W.
Higemoto, K. Nagamine, K. Matsuda, Y. Kohori, T. Kohara, Phys. B
\textbf{312-313}, 390 (2002).

22. \ N. Bernhoeft, G. H. Lander, M. J. Longfield, S. Langridge, D. Mannix, E.
Lidstroem, E. Colineau, A. Hiess, C. Vettier, F. Wastin, J. Rebizant, and P.
Lejay, to be published in the proceedings of SCES 2002, Krakow, Poland.

23. For a review, see ref. 7.

24. G. R. Stewart, Rev. Sci. Instrum. \textbf{54}, 1 (1983); B. Andraka, G.
Fraunberger, J.S. Kim, C. Quitmann, and G.R. Stewart, Phys. Rev. B39, 6420 (1989).

25. The differences between measuring specific heat in fields up to 33 T and
in the dc hybrid magnet up to 45 T, both at NHNMFL, were found to involve
fringe field effects on the measuring electronics due to the rather large
fringe field generated by the hybrid magnet. \ Orientation with respect to the
fringe field of the Keithley current sources was found to be important.

26. \ J. S. Kim and G. R. Stewart, unpublished.

27. B. Brandt, private communication.

28. K. Heuser, E.-W. Scheidt, T. Schreiner, Z. Fisk and G.R. Stewart, Journal
of Low Temp. Physics, \textbf{118,} 235 (2000)

29. Y. Imry and M. Wortis, Phys. Rev. B\textbf{19}, 3580 (1979).

30. \ The approximate field inhomogeneity expected for the hybrid magnet is 1
part in 10$^{4}$ for a position 5 mm from the field center. \ The single
crystal collage used for the hybrid magnet measurements were all quite thin (%
$<$
1mm) in vertical extent and collected on their sapphire disk within a 2 mm
radius from the center. \ A part in 10$^{4}$ in field would only be 0.0035 T
in a 35 T field, and the sample was well within 5 mm of field center in both
the radial and axial directions. \ Although Figs. 1 and 3 show that 0.3 to 0.4
T field changes significantly alter the specific heat anomaly near 34 and 38
T, there is no sign that a field inhomogeneity of order 0.0035 T has a
significant effect.

31. K. Sugiyama, M. Nakashima, H. Ohkuni, K. Kindo, Y. Haga, T. Honma, E.
Yamamoto, Y. Onuki, J. Phys. Soc. Japan \textbf{68}, 3394 (1999).

32. \ M. Jaime, K. H. Kim, G. Jorge, S. McCall, and J. A. Mydosh, cond-mat/0209500.
\end{document}